\documentclass[11pt,twoside]{article}

\usepackage{graphicx}
\usepackage{asp2006}
\usepackage{epsf}
\usepackage{lscape}

\begin{document}
\title{The Star-formation History of the Universe as Revealed from Deep Radio Observations of the 13$^H$ {\it XMM-Newton/Chandra} Deep Field}   
\author{{N. Seymour\altaffilmark{1}},  T. Dwelly\altaffilmark{2}, D. Moss\altaffilmark{2}, I. M$^{\rm c}$Hardy\altaffilmark{2}, M. Page\altaffilmark{3}, N. Loaring\altaffilmark{4,3} and G. Reieke\altaffilmark{5}}   

\altaffiltext{1}{{\it Spitzer} Science Center, California Institute of Technology, 1200 E. California Bd., Pasadena, CA, 91125, CA, USA.}
\altaffiltext{2}{School of Physics \& Astronomy, University of Southampton, Southampton, Hampshire SO17 1BJ, UK.}
\altaffiltext{3}{Mullard Space Science Laboratory, University College London, Holmbury St Mary, Dorking, Surrey RH5 6NT, UK.}
\altaffiltext{4}{SALT, PO Box 9, Observatory, 7935, South Africa.}
\altaffiltext{5}{Steward Observatory, University of Arizona, 933 North Cherry Avenue, Tucson, AZ 8572.}

\begin{abstract} 
  Discerning the exact nature of the faint (sub-mJy) radio population has been 
  historically difficult due to the low luminosity of these sources at most 
  wavelengths. Using 
  deep observations from {\it Chandra/XMM-Newton/Spitzer} and ground based 
  follow up we are able to disentangle the AGN and star-forming populations 
  for the first time in a deep multi-frequency GMRT/VLA/MERLIN Survey.
  The many diagnostics include radio luminosity, morphology, radio to mid-IR 
  flux density ratios, radio to optical flux density ratios and radio spectral 
  indices. Further diagnostics, e.g. optical spectra X-ray spectra/hardness 
  ratios, IR colours indicate the presence of the AGN {\em independent} of 
  whether the radio emission is powered by AGN or star-formation. We are able 
  to examine the star-formation history of the universe up to $z=2.5$ 
  in a unique way based on an unbiased star-formation rate indicator, radio 
  luminosity. This work provides an alternative perspective on the 
  distribution of star-formation by mass, ``downsizing'' 
  and allows us to examine the prevalence of AGN in star-bursts.
\end{abstract}

\section{Introduction}

Above a flux density of 1\,mJy the sources which dominate the 1.4\,GHz
radio source counts are almost exclusively AGN. However below 1mJy
there is an excess over the extrapolation from higher fluxes which is
known as the ``sub-mJy bump'' \citep{Seymour:Richards00, Seymour:Hopkins03,
  Seymour:Seymour04}.  The conventional view is that the bump arises from 
star-burst emission \citep{Seymour:Condon92, Seymour:Richards00}.  
Attempts to determine the separate evolution of the star-burst
and AGN populations have been severely hampered by the scarcity of
genuine multi-wavelength data sets on deep radio surveys. Source 
evolution is
constrained, rather poorly, only by the shape of the source counts
\citep[e.g. ][Moss et al. in prep.]{Seymour:Hopkins03, Seymour:Seymour04}

Only now with the advent of multi-wavelength data sets from X-ray 
to radio are we able to distinguish with confidence the nature of the 
radio emission for virtually all the sub-mJy radio population. This work 
is based on a deep radio surveys of the 13$^H$ {\it XMM-Newton/Chandra}
Deep Survey Field. In addition to $\sim16\,$days of MERLIN data, we 
have very deep observations at two frequencies: 1.4\,GHz down to 
rms$\sim7\,\mu$Jy from the VLA \citep{Seymour:Seymour04} and 610\,MHz down 
$\sim25\,\mu$Jy from the GMRT (Moss et al. in prep.). If we are able to 
distinguish between AGN and star-forming galaxies (SFGs) we will have 
an un-biased view of the star-formation history of the universe as radio
luminosity is a direct tracer of the star-formation rate (see Seymour et 
al. in prep).

\section{The Multi-wavelength Data Set}

We have a legion of complimentary data. Our deep multi-wavelength optical 
imaging from $0.3-2.2\,\mu$m allows us to derive photometric redshifts for 
the $\sim60\%$ of sources without spectroscopic redshifts (Dwelly et al. 
in prep). 

We also have deep X-ray data \citep{Seymour:McHardy03, Seymour:Loaring05, 
  Seymour:Page06} from {\it XMM-Newton} and {\it Chandra} as well deep 
observations from GTO {\it Spitzer} with the IRAC and MIPS instruments.

\section{Radio Emission and AGN Diagnostics}

The following diagnostics help us determine if the radio emission is due to 
AGN or star-formation activity:

{\bf a)} Radio luminosity: any sources with radio luminosities greater than 
$10^{25}$\,WHz$^{-1}$ are taken to be AGN.

{\bf b)} Radio morphology: objects with extended morphology, e.g. lobes and 
  jets, not associated with the underlying galaxy are taken to be AGN, but if 
  the morphology matches the underlying galaxy they are taken to be SFGs.

{\bf c)} Radio to mid-IR flux density ratio: any radio source with a $24\,\mu$m 
to 1.4\,GHz flux density ratio more than 0.3dex below that for the most 
extreme star-forming galaxies are taken to be AGN.

{\bf d)} Radio to optical flux density ratio: objects with radio to optical 
flux density ratios greater than 10 are taken to AGN. 

{\bf e)} Radio spectral index: any sources with radio spectral indices not in 
the range $-0.5 < \alpha < -1$ are taken to be AGN.

Indicators of the presence of AGN, independent of physical cause of the 
radio emission, include: optical spectra, mid-IR colours, X-ray 
luminosities and spectra.

\begin{figure}[!ht]
  \centering
  \includegraphics[width=2.8in,angle=270]{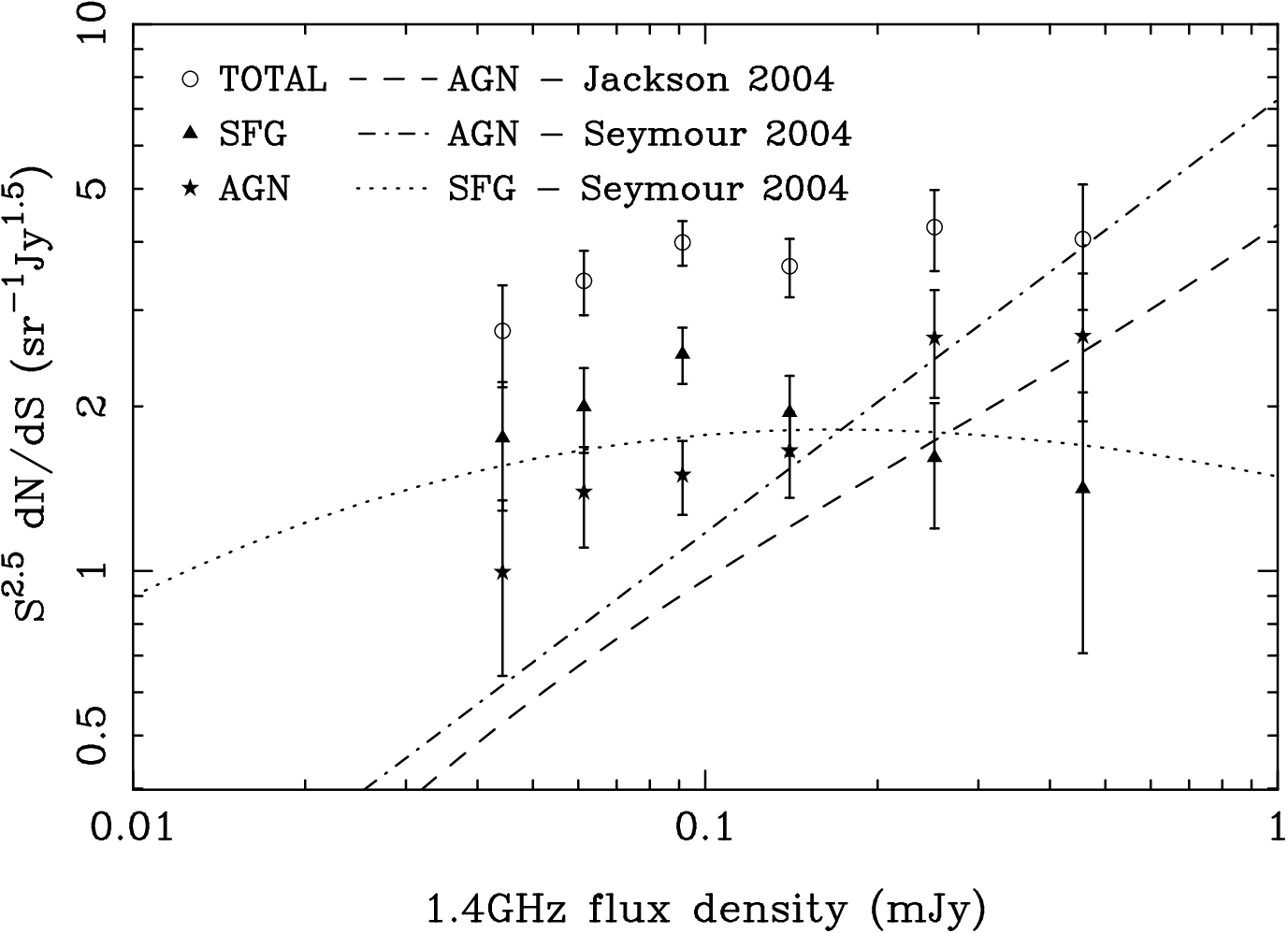}
  \caption{The 1.4\,GHz normalised Euclidean source counts. The total counts 
    are indicated by open circles, the contribution from radio AGN by stars
    and the contribution from radio star-forming galaxies by triangles. We
    overlay models for different components from \citet{Seymour:Seymour04} 
    based on models by \citet{Seymour:Hopkins98} and \citet{Seymour:Jackson04}.
    Part of the sub-mJy up-turn 
    appears to be due to a rise in the AGN contribution as well as that 
    of the star-forming population.}
  \label{seymour.fig.counts}
\end{figure}

\section{Radio Counts by Galaxy Type}

Using these radio emission diagnostics we find that $\sim43\%$ of the 
sources below $600\,\mu$Jy are AGN, $\sim52\%$ are SFGs and $\sim5\%$ are 
unknown. These $\sim5\%$ have no optical/near-IR counterparts and are assumed 
to assumed to be AGN at $z>2.5$. We are then 
able to re-derive the normalised Euclidean counts by galaxy type in 
Fig.~\ref{seymour.fig.counts}. The counts by type show the rapid increase 
in the star-forming population as predicted by \citet{Seymour:Hopkins98, 
Seymour:Seymour04} and references therein. However the contribution due to 
AGN does not decrease with flux density as rapidly as predicted by models 
extrapolated from higher flux densities and 
appears to contribute about $30\%$ of the faint counts below 0.1\,mJy.

\section{Star-formation History of the Universe Across $0\la z \la 2.5$}

\begin{figure}[!ht]
  \centering
  \includegraphics[width=3.2in,angle=270]{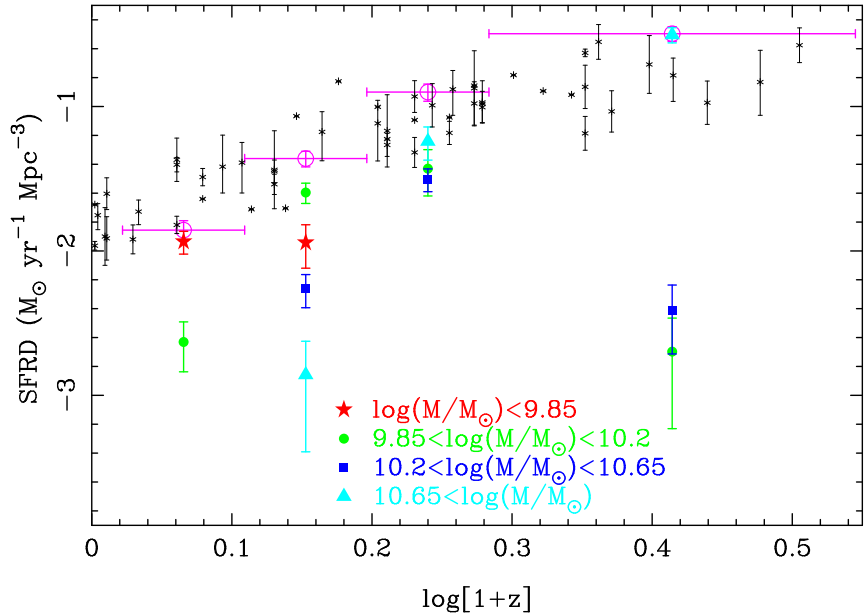}
  \caption{Here we present the star-formation rate density of the universe 
  across $0\la z \la 2.5$. The small black asterisks are from a compilation 
  by Hopkins et al. 2004.  The open circles are the total contribution from 
  the faint radio population where
  the vertical error bars represent Poisson statistics and the horizontal 
  ones the width of the bin in redshift. The stars, circles, squares and 
  triangles represent the contribution in each bin from galaxies of different 
  stellar mass with horizontal error bars omitted for clarity.}
  \label{seymour.fig.sfrd}
\end{figure}

With the successful separation of AGN and star-forming galaxies in our survey 
we are able to derive the star-formation rate density as a function of 
redshift \citep[e.g.][]{Seymour:Lilly96, Seymour:Madau96}. The star-formation 
rate (SFR) is derived from the radio luminosity using the relations of 
\citet{Seymour:Bell03} and then the SFR density as a function of redshift 
is determined by the $1/V_{max}$ 
method. The correction for the contribution to the total star-formation 
rate density in a given bin due to the radio sources below our detection 
limit is determined from our derivation of the evolution of the radio 
luminosity function (Dwelly et al. in prep.). We are further able to derive 
the contribution by stellar mass to the total star-formation rate density 
at each redshift using stellar masses determined from optical-near/mid-IR 
SED fitting, showing clear evidence of ``downsizing'' (see fig. 
\ref{seymour.fig.sfrd})
The fraction of AGN in the radio SFGs varies from $\sim20\%$ at low redshift 
to $\sim60\%$ in the highest redshift bin suggesting that AGNs may play a
crucial role in star-bursts at the peak redshift of star-formation, 
$1\la z \la2.5$.

\acknowledgements 
We would like to thank Jose Afonso and the LOC for organising a timely, 
interesting and thoroughly enjoyable meeting. 
We thank the staff of GMRT 
for their help in obtaining and reducing the data. We especially thank
Niruj Mohan for his help in all matters. 
We look forward to more Pages.

\end{document}